\documentclass{article}

\usepackage{arxiv}

\usepackage[utf8]{inputenc} 
\usepackage[T1]{fontenc}    
\usepackage{hyperref}       
\usepackage{url}            
\usepackage{booktabs}       
\usepackage{amsfonts}       
\usepackage{nicefrac}       
\usepackage{microtype}     
\usepackage{amsmath}
\usepackage{amssymb}
\usepackage{graphicx}
\usepackage{dcolumn}
\usepackage{bm}
\usepackage{xcolor}

\usepackage{multirow}
\usepackage{pbox}
\usepackage{cite} 
\usepackage{nameref}
\usepackage{url}  
\usepackage{ifthen}  
\usepackage{multicol}   
\usepackage[utf8]{inputenc} 

\usepackage{graphicx}
\usepackage[tight,footnotesize]{subfigure}
\usepackage[noend]{algorithmic}
\usepackage{algorithm}

\usepackage{amsmath}                     
\usepackage{amssymb}

\usepackage{tcolorbox}

\usepackage{authblk}

\title{Decoy Selection for Protein Structure Prediction Via Extreme Gradient Boosting and Ranking}

\author[1]{\small Nasrin Akhter}
\author[2,*]{\small Gopinath Chennupati}
\author[2]{\small Hristo Djidjev}
\author[1, 3, 4]{\small Amarda Shehu}

\affil[1]{Department of Computer Science, George Mason University, Fairfax, VA, 22030, USA}
\affil[2]{Information Sciences (CCS-3) Group, Los Alamos National Laboratory, Bikini At al Rd., 87545, Los Alamos, USA}
\affil[3]{Department of Bioengineering, George Mason University, Fairfax, VA, 22030, USA}
\affil[4]{School of Systems Biology, George Mason University, Manassas, VA, 20110, USA}
\affil[*]{Corresponding author\thanks{nakhter3@gmu.edu, gchennupati@lanl.gov, djidjev@lanl.gov, amarda@gmu.edu}: gchennupati@lanl.gov}

\begin{document}
\maketitle

\begin{abstract}
	
	\textbf{Background:} Identifying one or more biologically-active/native decoys from millions of non-native decoys is one of the major challenges in computational structural biology. The extreme lack of balance in positive and negative samples (native and non-native decoys) in a decoy set makes the problem even more complicated. Consensus methods show varied success in handling the challenge of decoy selection despite some issues associated with clustering large decoy sets and decoy sets that do not show much structural similarity. Recent investigations into energy landscape-based decoy selection approaches show promises. However, lack of generalization over varied test cases remains a bottleneck for these methods. 
	
	\textbf{Results:} We propose a novel decoy selection method, ML-Select, a machine learning framework that exploits the energy landscape associated with the structure space probed through a template-free decoy generation. The proposed method outperforms both clustering and energy ranking-based methods, all the while consistently offering better performance on varied test-cases. Moreover, ML-Select shows promising results even for the decoy sets consisting of mostly low-quality decoys.
	
	\textbf{Conclusions:} ML-Select is a useful method for decoy selection. This work suggests further research in finding more effective ways to adopt machine learning frameworks in achieving robust performance for decoy selection in template-free protein structure prediction.
	
\end{abstract}

\section{Background}
\label{sec:Introduction}

Protein molecules play a vital role in controlling the biological activities of a cell. There are a number of attempts in wet laboratories to determine biologically-active/native tertiary structures as a route to decoding protein function~\cite{maximova2016principles}. Technological advances have now made it possible to generate hundreds of thousands of tertiary structures for a given amino-acid sequence, known as decoys, in a few CPU hours~\cite{shehu2015review}. The multiplicity of decoys necessitates recognizing high-quality, near-native decoys among hundreds of thousand of decoys in an ensemble. Identifying these near-native decoys is a challenging problem in computational structural biology, and is known as decoy selection. 

Template-free methods, which generate low-energy tertiary structures in the absence of one or more structural templates from homogeneous sequences, have now become prominent. The most popular ones include Rosetta~\cite{leaver2011rosetta3} and Quark~\cite{xu2012ab}. To compute the low-energy structures, these methods employ stochastic optimization to find local minimum of a selected energy/scoring function. A well known fact is that energy bias often does not lead to tertiary structures that are close to the native. Therefore, identifying near-natives from a large ensemble of decoys remains an open problem~\cite{kryshtafovych2014assessment}.

Consequently, other decoy selection strategies gained momentum due to the weak role of energy in recognizing near-native conformations, which is reflected in Critical Assessment of protein Structure Prediction (CASP)~\cite{kryshtafovych2014assessment} series of community wide experiments. Clustering-based methods dominate the model quality assessment (MQA) performed in CASP. Clustering-based decoy selection methods work on the notion that decoys are randomly distributed around the native structure which a consensus method ought to reveal. The clustering-based decoy selection performs better when the ensemble consists of mostly good quality decoys. However, if the sampling of decoys in the decoy generation stage is sparse, resulting in many dissimilar decoys in an ensemble, consensus methods fail to recognize exceptionally good decoys~\cite{moult2014critical}. Moreover, the time complexity incurred in clustering a large decoy ensemble creates another bottleneck.

In addressing the above challenges in decoy selection, we propose an alternative approach that takes advantage of the consensus methods and a machine learning technique. As described in~\cite{bryngelson1995funnels}, protein energy landscape reveals important statistical information regarding the conformational organization and pathway. In this paper, we leverage the quantitative knowledge garnered from the energy landscape of a protein molecule in a machine learning framework to address the challenges in decoy selection. Supervised machine learning methods are gaining prominence in computational biology applications. These methods generate predictive models that learn subtle patterns from the data without making any prior assumptions~\cite{michalski2013machine}. One of the biggest challenges for these predictive models is to succeed even when the dataset is extremely imbalanced. Data imbalance is a common problem in computational biology and bioinformatics~\cite{zhao2008protein}. For instance, one of the benchmark proteins in our experiments contains only $0.005\%$ of positive instances (near-natives) among $58,491$ decoys. Even in such a sparse decoy set, the proposed method successfully identifies the near-natives. Our method works as follows: first, the method extracts local structures from the energy landscape probed through a template-free protein structure prediction method; next, a machine learning-based decoy selection method uses these local structures to finally select groups of good quality decoys. The method outperforms state-of-the-art decoy selection strategies in~\cite{akhter2018extraction}. 

\subsection{Related Work}
\label{sec:RelatedWork}

The diverse collection of decoy selection strategies can be categorized into single-model, multi-model, quasi-single, and machine learning (ML) methods. Single-model methods predict quality on a per-decoy basis~\cite{uziela2016proq2}, these are physics-based and/or knowledge-based. Physics-based methods employ different atomic interactions such as electrostatic, Van Der Waals interactions, hydrogen bonding~\cite{brooks1983charmm, cornell1996second, lazaridis1999discrimination}, whereas the knowledge-based scoring functions employ statistical analysis of known native structures~\cite{miyazawa1999empirical, mcconkey2003discrimination, simons1999improved}. Between these two methods, knowledge-based methods are known to be more successful in predicting high quality decoys~\cite{park1996energy, felts2002distinguishing}. 

Cluster-based methods work on the premise that the decoys are randomly distributed around the 'true' answer~\cite{lorenzen2007identification,estrada2010automatic}, which is not entirely valid due to the inherent bias associated with the template-free protein structure prediction methods used to generate the decoys. Apart from the huge time-complexity incurred by clustering a large decoy ensemble, the cluster-based methods often fail to identify good quality decoys (near-natives) for hard targets, which are more sparsely sampled~\cite{moult2014critical}. Despite the bottlenecks, cluster-based decoy selection strategies have been the most popular methods in the decoy selection literature. Quasi-single models combine the single-model and consensus methods. First, some high quality reference structures are selected, then the remaining decoys in the ensemble are compared with the reference structures~\cite{jing2016sorting}. These methods are shown to perform better~\cite{kryshtafovych2014assessment, he2013protein, pawlowski2016mqapsingle}. 

Recent investigations are employing machine learning (ML) methods for decoy selection~\cite{manavalan2014random,nguyen2014dl,hurtado2018deep}. For instance, work in~\cite{mirzaei2016purely} uses Support Vector Machine (SVM) and uses a statistical scoring function GOAP~\cite{zhou2011goap} to distinguish native decoys from the non-native ones. Decoy selection through machine learning are mostly single-model methods. These methods leverage structural features of proteins to assess decoy quality. Work in~\cite{akhter2019non} employs non-negative matrix factorization for selecting the best cluster of decoys and the the best decoy in the decoy set, which can be further extended to large scale using the the distributed implementations~\cite{chennupati2020distributed} of NMF.

Deep learning has also become a popular approach to address ML problems in bioinformatics~\cite{li2019deep}. Along with a variety of applications, such as DNA sequencing~\cite{li2018deepsimulator}, enzyme function prediction~\cite{li2017deepre}, de-novo prediction of membrane proteins~\cite{wang2018predmp}, protein contact map prediction~\cite{wang2017accurate}, and protein secondary structure prediction~\cite{wang2016protein}, deep learning has been successfully utilized for protein decoy selection as well. For instance, a deep belief network-based protein quality estimation (decoy selection) model DeepQA outperforms SVM-based methods and achieves state-of-the-art performance on the CASP dataset~\cite{cao2016deepqa}. Convolutional neural network-based models have also observed success in protein decoy selection~\cite{cao2016deepqa,sato2019protein,hou2019protein}.

In this paper, we prefer to investigate shallow models, which, unlike deep architectures, do not place such high demands on the size of the training dataset in relation to the number of parameters. As our ability to expediently generate or obtain structure data grows, deep learning will surely provide an interesting way forward that we plan to pursue in tandem with strategies to reduce the dimensionality of the loss function.

In this paper, we employ an ML technique to a multi-model method that exploits local structures extracted from an energy landscape~\cite{nussinov2014second}. The proposed ML-based multi-model method offers promising results in terms of higher true positives and lower false positives.

\section{Methods}
\label{sec:Methods}

First, we elaborate on the concept of energy landscape that forms the basis of our decoy selection method. 

\subsection{Energy Landscapes to Basins}

The energy landscape is an instance of a more general fitness landscape that comprises a set of points $X$, a neighborhood $\mathcal{N}(X)$ defined on $X$, a distance metric on $X$, and a fitness function $f: X \rightarrow \mathbb{R}_{\geq 0}$ that assigns a fitness to every point in $X$. Moreover, the points in $X$ secure neighbors via the neighborhood function. In the context of decoy selection, the points $x \in X$ represent decoy structures, and the fitness function often designates an energy function. Effectively, the energy landscape of decoy structures characterizes the mapping of  structures to their internal energy and provides important quantitative information about the structure space.

A protein energy landscape features an ensemble of structural states near or far from the native state and an extensive collection of intermediate states that shape the multi-modal and multi-dimensional nature of the landscape~\cite{nussinovwolynes14}. The concept of a basin is connected to a local/focal minimum. A focal minimum in a landscape is surrounded by a basin of attraction, which is the set of points on the landscape from which steepest descent/ascent converges to that focal optimum. Barriers separate basins and regulate transitions of a system between different structural states corresponding to basins in the landscape. 

Under the energy landscape treatment, the biologically-active/native state(s) can be determined by identifying corresponding basins, which requires one to extract the underlying organization of decoys to identify basins in the landscape. One approach to achieve this objective is to embed the decoys in a connectivity data structure and utilize energies to identify basins. Consider an $\Omega$ set of decoys. The $\Omega$ can be embedded in a nearest-neighbor graph (nn-graph) $G = (V, E)$~\cite{CazalsDreyfus17}. The vertex set $V$ is populated with the decoys, and the edge set $E$ is populated by inferring the neighborhood structure of the landscape. The distance between two structures is measured via root-mean-squared-deviation (RMSD) after each of the structures is superimposed over some reference structures (arbitrarily, chosen to be the first in the ensemble); the superimposition minimizes differences due to rigid-body motions. Each vertex $u \in V$ is connected to vertices $v \in V$ if $d(u, v) \leq \epsilon$, where $\epsilon$ is a user-defined parameter. If the landscape has been sampled sparsely and in a non-uniform way, there is a possibility of creating a disconnected graph from a small $\epsilon$ value. One way to prevent such scenario is to increase the $\epsilon$ while controlling the density of the resulting nn-graph via the number of nearest neighbors of $u$.

The local minima of the landscape can be detected by analyzing the nn-graph. A vertex $u \in V$ is a local minimum if $\forall v \in V \, f(u) \leq f(v)$, where $v \in N(u)$ ($N(u)$ denotes the neighborhood of $u$). The remaining vertices are then assigned to basins as follows. Each vertex $u$ is associated a negative gradient estimated by selecting the edge $(u, v)$ that maximizes the ratio $[f(u) - f(v)]/d(u,v)$. From each vertex $u$ that is not a local minimum, the negative gradient is followed (via the edge that maximizes the above ratio) until a local minimum is reached. Vertices that reach the same local minimum are assigned to the basin associated with that minimum.

\subsection{Basin Selection via Basin Ranking}
\label{sec:bs-br}
The basins, extracted from the energy landscape, can be useful in decoy selection. Works in~\cite{akhter2018extraction,akhter2019unsupervised} shows that simple, ranking-based basin selection strategies outperform a standard clustering-based decoy selection method in terms of purity (percentage of true positives, penalizes the selected basin by the extent of false positives found in that basin). Basins can be ranked as a combination of basin characteristics. For instance, basins can be ranked merely as size (S), as a combination of size and the energy (S+E) of the focal minimum of that basin. The size of basin is computed by the number decoys that belong to a basin. On the other hand, size and energy are used as conflicting objectives in a multi-objective, Pareto-based selection strategy. In a multi-objective optimization, solution A dominates solution B, if A is better than or equal to B for all optimization objectives, and for at least one objective, A is strictly better than B. In the context of basins, Pareto Rank (PR) of Basin A is the number of basins that dominate A. The Pareto Count (PC) of basin A is the number of basins that A dominates. Specifically,  basins can be ranked with their PR, or with PR and PC (PR+PC). Empirical studies conducted in~\cite{akhter2018extraction} demonstrate the superiority of the Pareto-based basin selection strategies over both cluster-based, size and energy-based decoy selection methods. 

Despite good performance, ranking-based decoy selection strategies are unable to perform consistently well over all test cases regardless of their difficulty levels. Neither S+E nor PR+PC can provide fair performance (less false positives and more true positives in the selected clusters/basins) over all or most of the test cases. One would prefer a decoy selection method that is able to provide reasonably good performance for all or most of the test cases regardless of difficulty level or heterogeneity in structural characteristics. This is the premise of the work presented in this paper.

\subsection{Decoy Selection via ML and Ranking}
Shortcomings of ranking-based basin selection strategies necessitate a new basin selection strategy. On that premise, we present a novel basin-based decoy selection method, referred to as \mbox{ML-Select}, that employs machine learning techniques. The method operates in two phases: the first phase captures $n$ pure basins; while the second phase purifies the selected $n$ basins and offers top $k$ purified basins as output. Both the phases involve fitting a regression model and a selection approach (ranking) based on the regression results. To generalize across all possible difficulty levels of proteins, we randomly select two proteins per difficulty level (easy, medium, hard) to train the models. Therefore, the performance of our models is independent of a test case and difficulty levels. We now describe the two phases of ML-Select in further detail. 

\subsubsection{Phase 1}
In this phase, ML-Select predicts the purity of basins and ranks them based on the predicted values. We use two kinds of attributes: {\em Pareto} and {\em graph}-based attributes as features to build the regression model. The Pareto-based features are PR and PC, computed from treating basin size and focal energy as two conflicting optimization objectives~\cite{akhter2018extraction}. We assign the ranks to each basin that are calculated based on the PR and PC values associated with the given basin. Specifically, each basin is assigned two ranks based on their PR and PC values, which serve as two different features. 

The graph-based feature, {\em number of connected components}, characterizes a spatial attribute of the graphical representation of basins. The extracted basins from the {\em nn-graph} (of all the decoys in the dataset) using the Structural Bioinformatics Library (SBL)~\cite{CazalsDreyfus17} are essentially bags of decoys. Estimating the spatial structure of these decoys in a specific basin is hard. Therefore, we consider the number of connected components as one of the features for ML-Select. 

\begin{figure*}[!t]
	\centering
	\includegraphics[width=0.2\textwidth]{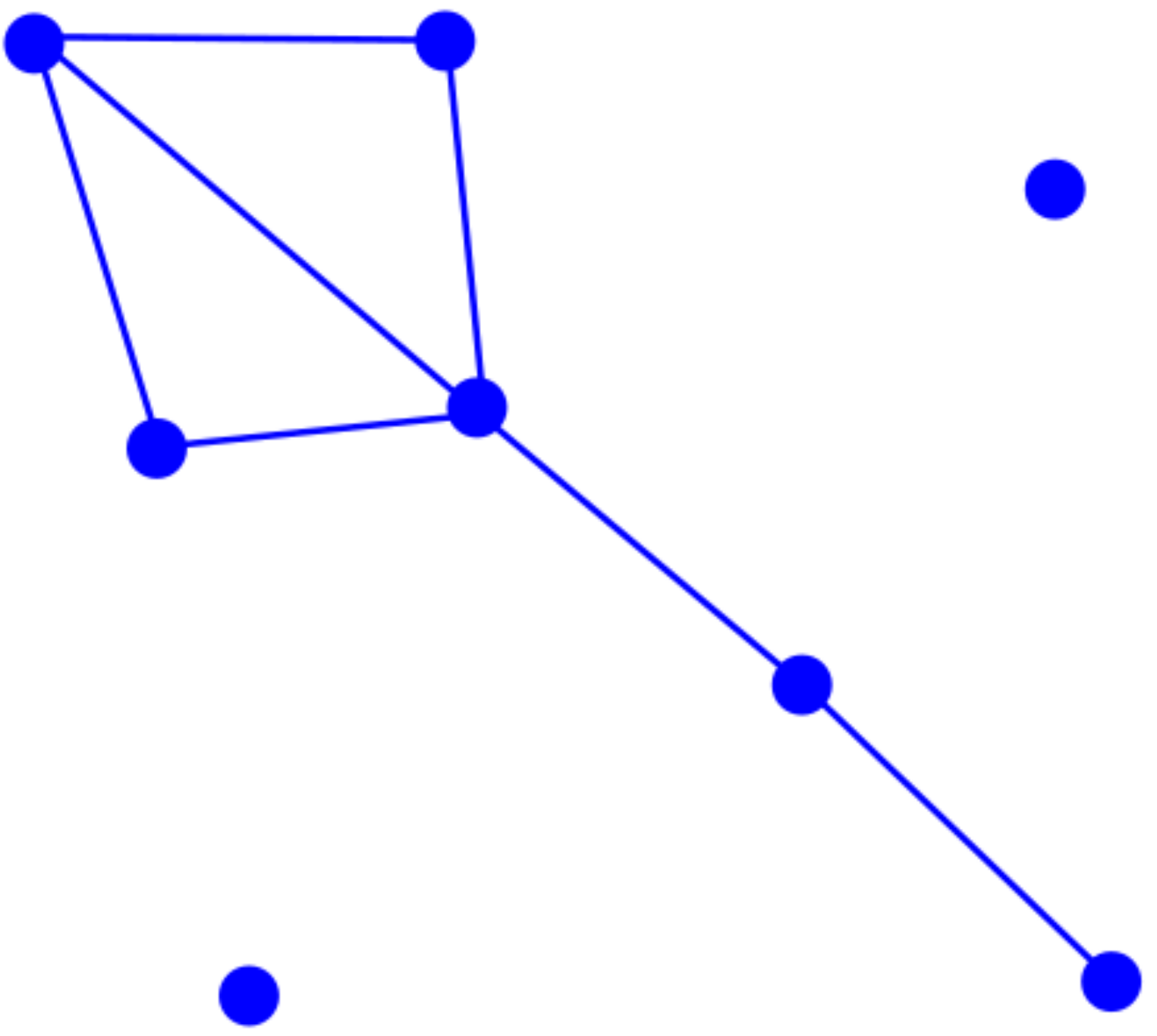}
	\caption{Three components in one of the basin-graphs of \textsl{1dtja}.}
	\label{fig:comp}
\end{figure*}

In order to easily recover the relative spatial organization of the decoys comprising a basin, we construct $m$ different nearest-neighbor graphs using the decoys populating $m$ different basins. We use $pdist+1$\AA \space for the distance threshold to create the nearest-neighbor graphs, where $pdist$ refers to the average pairwise distance between the decoys of the basins. Depending on the distance between the decoys in a basin, the corresponding graph may consist of one or more connected components, which signify the structural attribute of a basin. Figure~\ref{fig:comp} shows an example graphical representation of the components in a basin. We rank the basins based on the predicted purity and pass the top $n$ basins to the second phase for further purification. 

\subsubsection{Phase 2}
In the second phase, we predict the {\em root mean-squared-deviation} ($rmsd$) of a decoy from the true native. The training set of this phase uses the same proteins as in the first phase. However, the features in the second phase are different from that of the previous phase. We use twenty features of which three are knowledge-based potentials and the remaining are energy scores from Rosetta suite~\cite{RosettaScoring}. The three knowledge-based features are: $RW$, $RWplus$~\cite{zhou2002distance} and dDFIRE~\cite{yang2008specific}. $RW$ is distance-dependent atomic potential and $RWplus$ is side-chain orientation dependent potential; the third feature is $dDFIRE$, which improves the DFIRE statistical potential by adding an orientation dependency. The remaining $17$ features are energy terms in the {\em REF2015} scoring function~\cite{alford2017rosetta} in the {\em Rosetta} suite of scoring functions. 
The $17$ Rosetta REF2015 energy terms are the Lennard-Jones attractive and repulsive terms that capture interactions between atoms in different residues, the Lazaridis-Karplus solvation energy, the intra-residue Lazaridis-Karplus solvation energy term, the asymmetric solvation energy term, the Lennard-Jones repulsive term that captures interactions  between atoms in the same residue, the Coulombic electrostatic potential with a distance-dependent dielectric, the Proline ring closure energy and energy of the psi angle of preceding residue, the backbone-backbone hydrogen-bonding energy term between atoms close and distant in the primary sequence, the sidechain-backbone and sidechain-sidechain hydrogen-bonding energy term, the Ramachandran preferences term, the (backbone) omega dihedral term, the probability of amino acid given torsion values for the phi and psi backbonee angles, the internal energy of sidechain rotamers term (as derived from Dunbrack's statistics), and a special torsional potential term to keep the tyrosine hydroxyl in the plane of the aromatic ring.

The top $n$ pure basins from the first phase are treated as test cases. That is, we build $n$ regression models for $n$ basins that are passed to the second phase from the first phase. Each of these basins are further purified as follows. In a given basin from phase 1, if the predicted $rmsd$ of a decoy falls short of pre-defined threshold ($dist\_thresh$, explained later in the implementation details), we remove that decoy from a test case basin. Effectively, the decoys that are further away from the true native are removed from the selected basins. As a result, the purity of the selected basin improves. We rank the basins based on the resulting purity after the non-native decoy elimination and offer the top $k$ basins as a result at the end of second phase. The purification process in this phase poses a threat of eliminating a good decoy (ones near the native). We mitigate this effect with a shift in the pre-defined distance threshold, $dist\_thresh$ $\pm$ $\tau$, where $\tau$ $\in$ \{$10\%$, $20\%$, $25\%$\} of the pre-defined threshold. The effect of the threshold variation on purity is discussed later in the results. 

\subsection{Evaluation Metrics}
\label{sec:evaluation}
We evaluate the performance of our approach using two metrics: percentage of true positives ($n$) and purity ($p$).  At a given distance threshold $dist\_thresh$ (explained in the implementation details), $n$ is the ratio of number of true near-natives in the selected basin $B_{1-x}$, where $x \in \{1,2,3\}$, to the total number of true near-natives in that decoy ensemble. This metric resembles the Sensitivity (recall or true positive rate) measure. However, even significantly high $n$ might become less effective if the number of false positives in the selected basin is high, where, a random draw from the selected basin would result in a lower probability of offering a true near-native. The metric $p$ compensates this scenario by penalizing a large basin (or a group of selected basins) containing a large number of {\em true} and {\em false} positives to the extent of the false positive population present in that basin. $p$ is computed as a ratio of the number of true positives to the size of a basin (or a group of basins). Therefore, a basin with a large number of false positives results in a low purity regardless of the number of true positives in that basin. In essence, purity metric resembles the precision of our method. Specifically, we discuss the performance of ML-Select and four other competing methods in terms of purity metric due to its balanced treatment towards false and true positives. For evaluation, we select these metrics that focus more on true and false positives rather than on true and false negatives because here we are more concerned with increasing the probability of selecting a true positive from the selected basins in a random draw, which can be achieved by minimizing the false positives and maximizing the true positives. 

\subsection{Implementation Details}
\label{sec:implementation}
We use a distance threshold of 1\AA~for creating the nn-graph of a decoy ensemble via SBL~\cite{CazalsDreyfus17}. Since Rosetta decoy generation protocol may produce sparse samples, a low threshold may result in a disconnected graph. To address this problem, we increase the initial threshold until the graph is connected. Minimum distance from a decoy in an ensemble to the true native is referred to as $min\_dist$. For a protein with a known native structure, all decoys under the threshold $dist\_thresh$ are deemed as near-natives. As there are three different categories of test cases, we set the $dist\_thresh$ parameter to determine the near-natives on a per-case basis. More specifically, $dist\_thresh$ is set to 2\AA \space for the easy cases ($min\_dist < 1\AA$). For the medium cases ($1\AA \leq min\_dist < 2\AA$), $dist\_thresh$ is either $2.5\AA$ or $3\AA$. For the hard cases ($3\AA < min\_dist$), we increase the $dist\_thresh$ until one of the methods accumulate non-zero number of near-natives in the top selected basins. Moreover, if any test case belongs to a particular category based on the $min\_dist$, but very few near-natives can be found according to that $min\_dist$, we move that test case to the next difficulty level. 

We use a boosting-based ensemble learning approach, XGBoost~\cite{friedman2001greedy}, to build the regression models. We use a linear regression model via XGBoost in both phase 1 and phase 2. XGBoost is fast, scalable that follows the principle of gradient boosting. XGBoost is good to control over-fitting while producing a more regularized model formalization~\cite{chen2016xgboost}. We calculate the knowledge-based features as follows. We calculate the \textsl{RW} potentials in the form of \textsl{calRW} and \textsl{calRWplus}, the executable programs used in the calculation are from Zhang lab~\cite{RWpotential}. The dDFIRE potential has been calculated using dDFIRE program~\cite{dDFIREpotential}. We use $15$ rounds of boosting to build our regression model. For training the regression models, we choose top $q$ pure basins and randomly draw $q$ basins (total $2q$ basins) from the rest of the training data. 

We use $2$ easy, $2$ medium, and $2$ hard proteins for training the models. For testing, we use an easy, a medium, or a hard protein that has not been used in the training dataset. To test/evaluate on a protein, we use another protein to take its place for training. Eventually, all the 18 proteins are tested and there is no overlap between the training and testing data. 

To address the randomness in the training phase, we run the models on the test data for $50$ times, and report the average $p$ and $n$. We use $10$ for $q$ in this experiment. Construction of the nn-graph by SBL takes from $1$ to $2$ hours depending on the lengths (number of amino acids) of the proteins and the size of the decoy ensembles. Construction of the regression models take about a minute. Once the model has been built, testing it on a new dataset with $50$ runs takes about 12 seconds. Basin-Size and Basin-Size+Energy take about 20 seconds to test a new dataset. The runtimes for Pareto-Rank and Pareto-Rank+Count are $65$ and $96$ seconds, respectively.

\begin{table}[htp] \footnotesize
	\caption{Testing dataset (* denotes proteins with a predominant
		$\beta$ fold and a short helix).}
	\centering
	\begin{tabular}{|p{4.2em}|p{0.7em}|p{3.8em}|p{2.9em}|p{3.0em}|p{3em}|p{3.7em}|} \hline
		\multirow{2}{*}{\textbf{Difficulty}} & \multirow{2}{*}{\#} & \multirow{2}{*}{\textbf{PDB ID}} & \multirow{2}{*}{\textbf{Fold}} & \multirow{2}{*}{\textbf{Length}} & \multirow{2}{*}{\textbf{$|\Omega|$}} & \textbf{min\_dist} \\
		& & & & & & (\AA)\\
		\hline
		\multirow{5}{*}{Easy}  
		& $1$ & 1dtdb   & $\alpha+\beta$       & $61$   & $58,745$   & $0.51$\\ 
		& $2$ & 1wapa  & $\beta$            & $68$     & $68,000$    & $0.68$ \\
		& $3$ & 1hz6a   & $\alpha+\beta$   & $64$   & $60,000$   & $0.69$\\
		& $4$ & 1tig   & $\alpha+\beta$   & $88$   & $60,000$   & $0.70$\\
		& $5$ & 1dtja  & $\alpha+\beta$   & $74$   & $60,500$   & $0.74$\\ 
		\hline
		\multirow{7}{*}{Medium}
		& $6$ & 1bq9  & $\beta$ 		& $53$   & $61,000$   & $0.98$ \\ 
		& $7 $ & 1ail  & $\alpha$        & $70$   & $58,491$   & $1.01$ \\
		& $8 $ & 1c8ca   & $\beta^*$   	& $64$   & $65,000$   & $1.04$ \\
		& $9 $ & 2ci2   & $\alpha+\beta$   & $65$   & $60,000$   & $1.19$ \\
		& $10$ & 1fwp   & $\alpha+\beta$ & $69$   & $51,724$   & $1.63$\\
		& $12$ & 1sap   & $\beta$        & $66$   & $66,000$   & $1.93$ \\ \hline
		\multirow{6}{*}{Hard}
		& $11$ & 1hhp   & $\beta^*$  	& $99$  & $60,000$   & $1.85$ \\
		& $13$ & 2ezk  & $\alpha$       & $93$  & $54,626$   & $2.89$ \\
		& $14$ & 1aoy   & $\alpha$       & $78$   & $57,000$   & $3.03$\\
		& $15$ & 2h5nd   & $\alpha$       & $123$   & $54,795$   & $3.46$ \\
		& $16$ & 1isua   & $coil$       & $62$   & $60,000$   & $3.67$\\
		& $17$ & 1cc5  & $\alpha$       & $83$   & $55,000$   & $4.31$\\
		& $18$ & 1aly   & $\beta$        & $146$  & $53,000$   & $9.38$\\ \hline
	\end{tabular}
	\label{table:dataset} 
\end{table}

\section{Results}
\label{sec:Results}
We experimented with eighteen proteins of different lengths and folds. These proteins constitute a benchmark dataset often used by decoy generation algorithms~\cite{MeilerBaker2003, Bartolo10, OlsonShehuBCB13, MolloyShehuTCBB13, ZhangYu17, ZhangZhou18}. We used the Rosetta \textit{template-free} (decoy generation) protocol to generate around $51,000$ to $68,000$ decoys per target. Table~\ref{table:dataset} presents all the eighteen proteins arranged into three different categories (easy, medium, and hard). The difficulty level (easy, medium, hard) has been determined using the minimum distance ($min\_dist$) between the generated decoys and a known native conformation of the corresponding protein. The size of the decoy ensemble $|\Omega|$ for each target is shown in column $6$.

\begin{figure*}[!t]
	\centering
	\includegraphics[width=0.94\textwidth,height=1.28\textwidth]{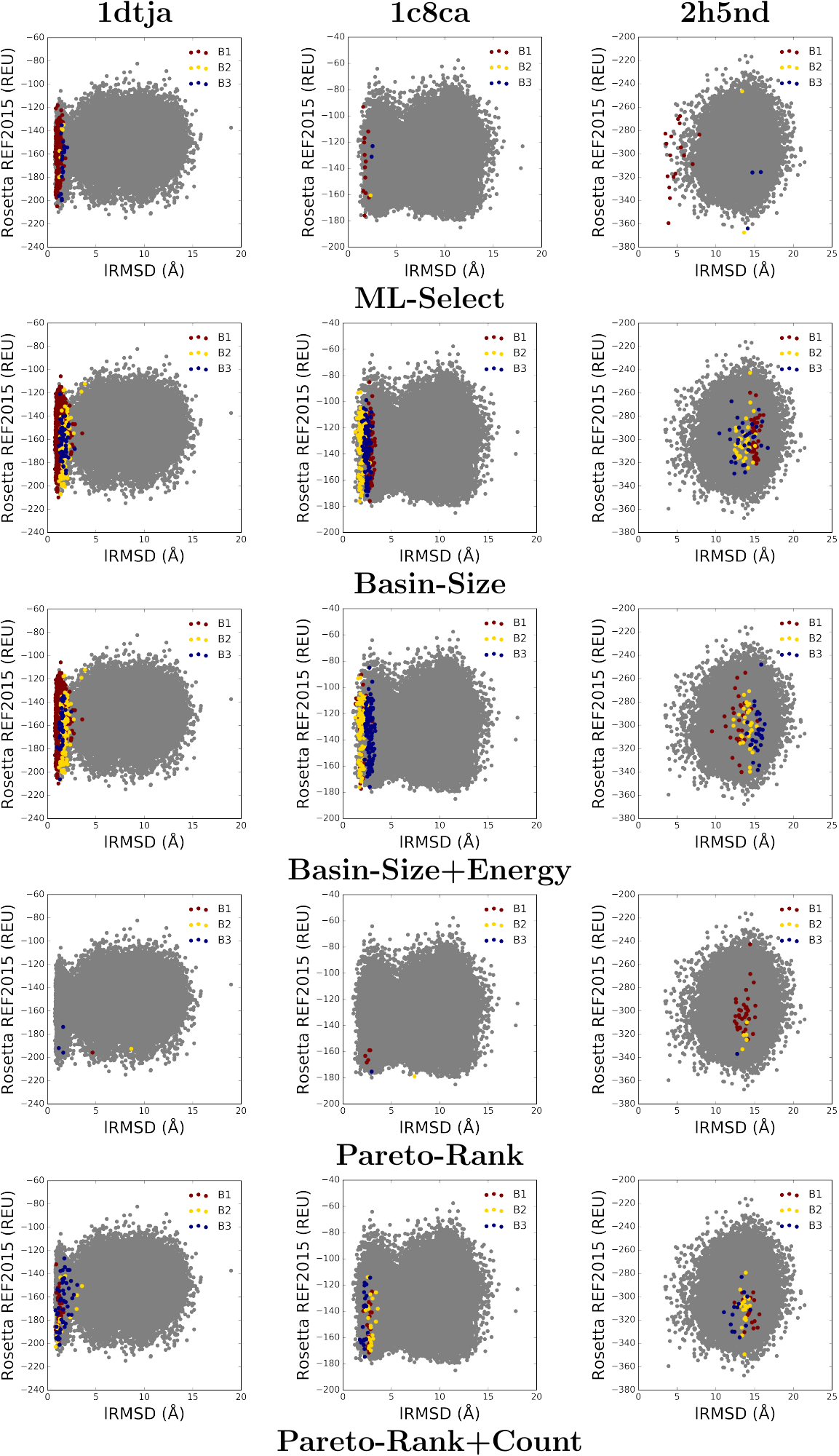}
	\caption{Visualization of selected decoys for three target proteins (indicated by the PDB id of their native structure). Decoys are plotted by their lRMSD from the native structure and their Rosetta REF2015 all-atom energy.}
	\label{fig:basins}
\end{figure*}

\subsection{Visualizing Top Basins}
\label{sec:VisualComparison}
Figure~\ref{fig:basins} provides a visual comparison of the methods with respect to the quality of the selected decoys in the top three basins. We present three representative cases from the easy, medium, and hard categories. Each plot shows the decoys as two-dimensional dots where the x-axis tracks the lRMSD of each decoy and the y-axis tracks the Rosetta REF2015 (all-atom) energy (measured in Rosetta Energy Units - REUs). Decoys in each basin are colored in maroon, gold, and navy to distinguish between the top three basins.

The protein with known native structure under PDB id \textsl{1dtja}, shown in the first column in Figure~\ref{fig:basins}, presents an easy case. ML-Select, shown in top row, captures the best quality decoys (near-natives, low lRMSD from the native) in the top three basins ($p: 99.6\%$). All the decoys in top three basins are within $2$\AA \space from the known native. On the other hand, the top three basins, selected by four other strategies, contain decoys with larger lRMSD, which lowers the purity (as low as $60\%$). For instance, Pareto-Rank captures very few decoys in top three basins. Moreover, some of these decoys are more than $4$\AA \space away from the native.

Although ML-Select obtains basins of smaller size compared to that of the existing strategies for the medium case, \textsl{1c8ca}, the quality of the selected decoys are better, which results in higher purity ($100\%$, $99\%$, $89.1\%$ for $B_1$, $B_{1-2}$, $B_{1-3}$, respectively). Contrarily, the larger basins, selected by Basin-Size, PR, and PR+PC, suffer from low purity due to the presence of numerous non near-natives (minimum $4.9\%$ and maximum $52.7\%$). Basin-Size+Energy performs fair in this scenario ($p: 94.4\%$ for $B_{1-2}$). However, purity diminishes as more basins are added in the selection ($56.2\%$ for $B_{1-3}$). Evidently, it is more likely that a random draw would yield a near-native from the top basin (or group of basins) if ML-Select is employed to perform the selection.

ML-Select excels even in the hard cases, as shown for the protein with known native structure under PDB id \textsl{2h5nd}. The quality of the decoys selected in ML-Select is as good as the Rosetta structure prediction protocol can sample ($p: 94.1\%$ for $B_1$). None of the existing basin-based strategies provide any near-native in their selected basins. That is, all the top basins selected by four other decoy selection strategies contain only false positives (decoys with larger lRMSD from the native ($\geq 10$\AA)).

\begin{figure*}[!t]
	\centering
	\includegraphics[width=0.55\textwidth]{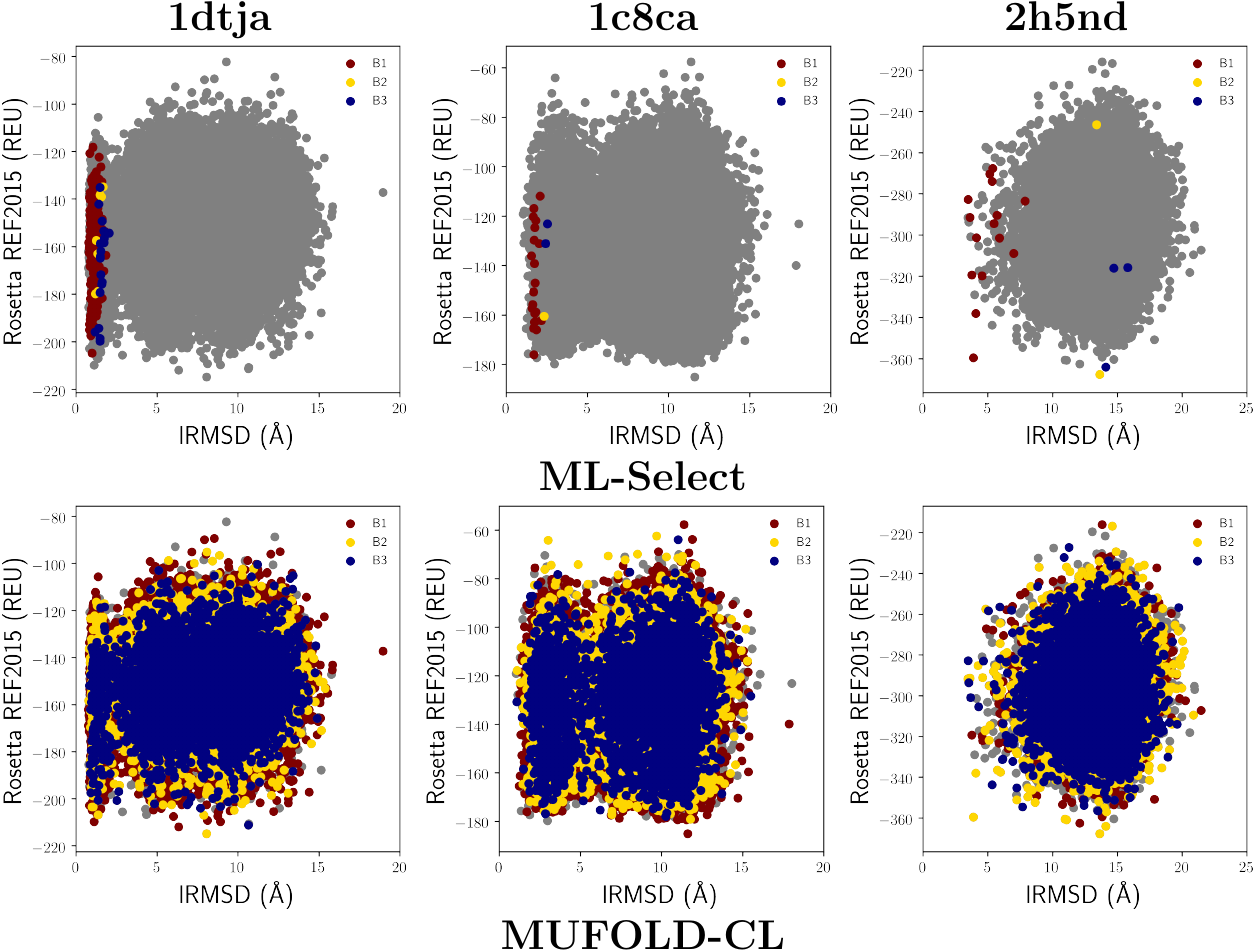}
	\caption{Visualization of decoys selected by ML-Select and MUFOLD-CL for three target proteins (indicated by the PDB id of their native structure). Decoys are plotted by their lRMSD from the native structure and their Rosetta REF2015 all-atom energy.}
	\label{fig:basins_with_mufold}
\end{figure*}

Figure~\ref{fig:basins_with_mufold} compares the top $3$ basins selected by ML-Select with the top $3$ clusters selected by a state-of-the art clustering-based model quality estimation method, MUFOLD-CL~\cite{zhang2013fast}. Since larger clusters are considered to have tighter distributions and are typically used for near-native model selection in practice~\cite{zhang2013fast}, we select the three largest clusters resulting from MUFOLD-CL as the top three clusters for comparison. As shown in Figure~\ref{fig:basins_with_mufold}, the top three clusters resulting from MUFOLD-CL are much larger; they contain near-natives, as well as as many non-natives. The presence of many non-natives lowers purity. For instance, for the easy protein \textsl{1dtja}, despite containing $57.3\%$ near-natives in the top cluster, purity is only $3\%$. This is due to the presence of many non-natives.

\begin{table*}[htp]
	\tiny
	\centering
	\caption{Comparison of the five basin-selection strategies. The
		top $G_{1-x}$ groups of decoys selected from each selection
		strategy, with $x$ limited to $3$, are analyzed. When analyzing $B_{1-x}$, the top $x$ basins are
		merged. The analysis lists the metrics (M): percentage of near-native decoys
		($n$); the purity ($p$), which is the proportion of near-native
		decoys relative to the size of a group; and the relative size ($s$, is proportional to $|\Omega|$) of each basin.}
	\begin{tabular}{|l|p{.3em}|p{2.2em}|p{2.2em}|p{2.2em}||p{2.2em}|p{2.2em}|p{2.2em}||p{2.2em}|p{2.2em}|p{2.2em}||p{2.2em}|p{2.2em}|p{2.2em}||p{2.2em}|p{2.2em}|p{2.2em}|}
		\toprule
		
		\multirow{2}{*}{\#} & \multirow{2}{*}{\textbf{M}} & \multicolumn{3}{c|}{\textbf{ML-Select}} &  \multicolumn{3}{c|}{\textbf{Basin-Size}} & \multicolumn{3}{c|}{\textbf{Basin-Size+Energy}} & \multicolumn{3}{c|}{\textbf{PR}} & \multicolumn{3}{c|}{\textbf{PR+PC}}\\
		\cline{3-17}
		& & B$_1$  & B$_{1-2}$ & B$_{1-3}$ & B$_1$ & B$_{1-2}$ & B$_{1-3}$ & B$_1$ & B$_{1-2}$ & B$_{1-3}$ & B$_1$ & B$_{1-2}$ & B$_{1-3}$ & B$_1$ & B$_{1-2}$ & B$_{1-3}$\\
		\hline
		
		\multirow{3}{*}{\hspace*{-1mm}1dtdb} & n
		& 11.2\% & 11.3\% & 11.7\%  & 88.3\% & 92.4\% & 92.4\% & 88.3\% & 92.4\% & 92.4\% & 0\% & 0\% & 0\% & 4.1\% & 5.1\% & 93.4\% \\
		& p & \textbf{100\%} & \textbf{100\%} & \textbf{100\%}  & 99.6\% & 99.6\% & 99.9\% & 99.6\% & 99.6\% & 97.2\% & 0\% & 0\% & 0\% & \textbf{100\%} & \textbf{100\%} & 93.4\% \\
		& s & 0.26\%  & 0.26\% & 0.3\%  &  2.1\% & 2.2\%  & 2.2\% & 2.1\% & 2.2\%  & 2.2\% & 0.002\% & 0.009\% & 0.01\% & 0.09\% & 0.12\% & 2.2\% \\
		\hline\hline
		
		\multirow{3}{*}{\hspace*{-1mm}1wapa} & n
		& 0.31\% & 0.6\% & 0.8\%  &  83.7\% & 83.7\%   & 83.7\% & 83.7\% & 83.7\%   & 83.7\% & 0\% & 0\% & 2.3\% & 0\% & 0\% & 0\% \\
		
		& p & \textbf{100\%} & \textbf{100\%}   & \textbf{100\%}  & 99\%   & 87.8\% & 79.3\% & 99\% & 89.4\% & 81.5\% & 0\% & 0\% & 80\% & 0\% & 0\% & 0\% \\
		
		& s& 0.002\% & 0.003\% & 0.004\%  & 0.43\% & 0.48\% & 0.5\% & 0.43\% & 0.47\% & 0.5\% & 0.001\% & 0.003\% & 0.02\% & 0.02\% & 0.05\% & 0.08\%\\
		\hline\hline
		
		\multirow{3}{*}{\hspace*{-1mm}1hz6a} & n
		& 4.6\% & 4.6\% & 4.6\%  &  35.9\% & 35.9\%   & 44.7\% & 35.9\% & 35.9\% & 35.9\% & 0.02\% & 0.02\% & 0.02\% & 2.2\% & 2.6\% & 4.6\% \\
		
		& p & 99.8\% & \textbf{99.4\%}   & \textbf{98.5\%}  &  99.7\%   & 77.6\% & 81\% & 99.7\% & 92.3\% & 73\% & 67.6\% & 67.1\% & 66.2\% & \textbf{100\%} & 99.3\% & 92.2\%\\
		& s &0.44\% & 0.44\% & 0.45\%  & 3.4\%   & 4.4\%  & 5.3\% & 3.4\% & 3.7\% & 4.7\% & 0.25\% & 0.25\% & 0.25\% & 0.21\% & 0.25\% & 0.47\% \\
		\hline\hline
		
		\multirow{3}{*}{\hspace*{-1mm}1tig} & n
		& 3.7\% & 6.2\% & 7.1\%  & 35.2\% & 42.9\%   & 48\% & 35.2\% & 42.9\%   & 48\% & 0\% & 0\% & 0\% & 2.9\% & 4.1\% & 5.6\% \\
		
		& p & \textbf{100\%} & \textbf{100\%}   & \textbf{100\%}  & 99.6\%   & 99.5\% & 99.1\% & 99.6\%   & 99.5\% & 99.1\% & 0\% & 0\% & 0\% & \textbf{100\%} & \textbf{100\%} & 98.7\% \\
		& s & 0.08\%  & 0.14\%  & 0.16\%  &  0.8\%  & 0.95\%  & 1.1\% & 0.8\%  & 0.95\%  & 1.1\% & 0.002\% & 0.003\% & 0.005\% & 0.06\% & 0.09\% & 0.13\% \\
		\hline\hline
		
		\multirow{3}{*}{\hspace*{-1mm}1dtja} & n
		& 7.5\% & 7.9\% & 8.6\%  & 54.5\% & 59.4\%   & 62\% & 54.5\% & 59.4\%   & 61.4\% & 0\% & 0\% & 0.15\% & 2.15\% & 2.84\% & 4.94\% \\
		
		& p & \textbf{100\%} & \textbf{100\%}   & \textbf{99.6\%}  & 98.6\%   & 97.3\% & 97\% & 98.6\%   & 97.3\% & 97.3\% & 0\% & 0\% & 60\% & \textbf{100\%} & 93.1\% & 87\% \\
		& s & 0.23\% & 0.25\%   & 0.27\%  &  1.74\%    & 1.9\% & 2\% & 1.74\%    & 1.9\% & 1.98\%  & 0.002\%  & 0.003\%  & 0.008\%  & 0.07\%  & 0.09\%  & 0.18\%  \\
		\hline\hline
		
		\multirow{3}{*}{\hspace*{-1mm}1bq9} & n
		& 0.62\% & 1.4\% & 2.4\%  &  0\% & 0\%   & 0\% & 0\% & 0\%   & 0\% & 0\% & 0\%   & 0\% & 0\% & 0\%   & 0\% \\
		
		& p & \textbf{100\%} & \textbf{95.1\%}   & \textbf{83\%}  & 0\%   & 0\% & 0\% & 0\%   & 0\% & 0\%  & 0\%   & 0\% & 0\%  & 0\%   & 0\% & 0\%  \\
		
		& s & 0.002\% & 0.004\%  & 0.01\%   &  0.07\%    & 0.12\% & 0.17\%  & 0.04\%     & 0.1\%     & 0.16\%     & 0.002\%     & 0.02\%     & 0.05\%     & 0.02\%     & 0.04\%     & 0.06\%    \\
		\hline\hline
		
		\multirow{3}{*}{\hspace*{-1mm}1ail} & n 
		& 1.4\% & 3.8\%  & 3.8\%  & 0\% & 0\%   & 0\% & 0\% & 0\%   & 0.92\%  & 0\%  & 0\%  & 0\%  & 0\%  & 0\%  & 0.3\%  \\
		
		& p & \textbf{100\%} & \textbf{92.5\%}   & \textbf{86\%}  & 0\%   & 0\% & 0\% & 0\%   & 0\% & 3\%  & 0\%  & 0\%  & 0\%  & 0\%  & 0\%  & 1.6\%  \\
		
		& s & 0.01\% & 0.023\%    & 0.025\%  &  0.14\% & 0.22\% & 0.3\% & 0.05\%  & 0.13\%  & 0.17\%  & 0.001\%  & 0.005\%  & 0.008\%  & 0.034\%  & 0.063\%  & 0.11\%  \\
		\hline\hline
		
		\multirow{3}{*}{\hspace*{-1mm}1c8ca} & n 
		& 0.8\% & 1.0\%     & 1.1\%    & 1.1\%     & 6.2\%   & 8.6\% & 1.4\%   & 6.5\%   & 7.6\%   & 0.11\%   & 0.11\%   & 0.11\%   & 0.06\%   & 0.11\%   & 1.21\%   \\
		& p & \textbf{100\%} & \textbf{99\%} & \textbf{89.1\%}    & 16.7\%   & 52.1\%  & 52.7\% & 86.2\%  & 94.4\%  & 56.2\%  & 40\%  & 33.3\%  & 28.6\%  & 5.3\%  & 4.9\%  & 34.9\%  \\
		
		& s & 0.02\% & 0.03\% & 0.034\%  &  0.18\%    & 0.33\%  & 0.46\% & 0.044\%  & 0.2\%  & 0.4\%  & 0.009\%  & 0.009\%  & 0.01\%  & 0.03\%  & 0.06\%  & 0.1\%  \\
		\hline\hline
		
		\multirow{3}{*}{\hspace*{-1mm}2ci2} & n
		& 0\% & 0\%   & 0\%   & 0\%   & 0\%   & 0\% & 0.77\%  & 0.77\%  & 0.77\%  & 0.51\%  & 0.51\%  & 0.51\%  & 0\%  & 0\%  & 0.26\%  \\
		
		& p & 0\% & 0\%   & 0\%  &  0\%   & 0\% & 0\% & 48.4\%  & 23.4\%  & 15.2\%  & \textbf{90.9\%}  & \textbf{83.3\%}  & \textbf{76.9\%}  & 0\%  & 0\%  & 7.6\%  \\
		
		& s & 0.01\%  & 0.02\%    & 0.03\%   &  0.06\%   & 0.12\%   & 0.18\% & 0.05\%  & 0.1\%  & 0.17\%  & 0.02\%  & 0.02\%  & 0.021\%  & 0.03\%  & 0.07\%  & 0.11\%  \\
		\hline\hline
		
		\multirow{3}{*}{\hspace*{-1mm}1fwp} & n 
		& 1.84\%    & 4.5\%  & 4.5\%  &  0\%   & 0\%   & 0\% & 0\%  & 0\%  & 0\%  & 9.3\%  & 9.3\%  & 9.3\%  & 0\%  & 1.3\%  & 1.3\%  \\
		
		& p & \textbf{97.7\%}    & \textbf{75.4\%}   & 60.3\%   &  0\%   & 0\%     & 0\% & 0\%   & 0\%   & 0\%   & 77.8\%   & 70\%   & \textbf{63.6\%}   & 0\%   & 3.7\%   & 2.4\%   \\
		
		& s & 0.003\% & 0.008\%  & 0.01\%  &  0.06\%    & 0.12\%    & 0.17\% & 0.05\%  & 0.1\%  & 0.15\%  & 0.017\%  & 0.019\%  & 0.02\%  & 0.03\%  & 0.05\%  & 0.08\%  \\
		\hline\hline
		
		\multirow{3}{*}{\hspace*{-1mm}1sap} & n 
		& 2.63\%   & 2.63\%   & 2.63\%    & 9.3\%    & 14.8\%   & 20.9\% & 0\%  & 1.5\%  & 10.8\%  & 0\%  & 0\%  & 0\%  & 0.4\%  & 0.8\%  & 1.6\%  \\
		
		& p & 87.8\%   & 71.7\%   & 70.6\%    &  85\%      & 84.6\%  & 88.3\% & 0\%  & 26.9\%  & 65.4\%  & 0\%  & 0\%  & 0\%  & \textbf{100\%}  & \textbf{86.7\%}  & \textbf{92.4\%}  \\
		
		& s & 0.21\% & 0.25\%  & 0.26\%     &  0.8\%    & 1.2\%   & 1.7\% & 0.2\%  & 0.4\%  & 1.2\%  & 0.002\%  & 0.003\%  & 0.005\%  & 0.03\%  & 0.07\%  & 0.12\%  \\
		\hline\hline
		
		\multirow{3}{*}{\hspace*{-1mm}1hhp} & n
		& 12.2\%   & 18.3\%   & 24.2\%  &  0\%   & 0\%   & 0\% & 0\%  & 0\%  & 0\%  & 0\%  & 0\%  & 0\%  & 0\%  & 0\%  & 0\%  \\
		
		& p & \textbf{84.2\%}   & \textbf{74.8\%}   & \textbf{68\%}   &  0\%   & 0\%   & 0\% & 0\%  & 0\%  & 0\%  & 0\%  & 0\%  & 0\%  & 0\%  & 0\%  & 0\%  \\
		
		& s & 0.012\% & 0.02\% & 0.03\%  &  0.06\% & 0.13\% & 0.19\% & 0.06\% & 0.1\% & 0.16\% & 0.007\% & 0.03\% & 0.08\% & 0.03\% & 0.06\% & 0.08\% \\
		\hline\hline
		
		\multirow{3}{*}{\hspace*{-1mm}2ezk} & n 
		& 1.3\%   & 1.3\%   & 1.3\%    &  0\%    & 0\%   & 0\% & 1.83\%  & 1.83\%  & 1.83\%  & 0\%  & 0\%  & 0\%  & 0\%  & 0\%  & 0\%  \\
		
		& p & \textbf{59.3\%}   & \textbf{45.6\%}   & \textbf{40.3\%}    &  0\%      & 0\%  & 0\% & 51.6\%  & 19.8\%  & 14\%  & 0\%  & 0\%  & 0\%  & 0\%  & 0\%  & 0\%  \\
		
		& s & 0.03\% & 0.045\%  & 0.51\%     &  0.09\%    & 0.16\%   & 0.23\% & 0.06\%  & 0.15\%  & 0.21\%  & 0.01\%  & 0.02\%  & 0.03\%  & 0.03\%  & 0.07\%  & 0.11\%  \\
		\hline\hline
		
		\multirow{3}{*}{\hspace*{-1mm}1aoy} & n
		& 0.11\%   & 0.23\%   & 0.29\%    & 0.12\%    & 0.12\%   & 0.15\% & 0.03\%  & 0.2\%  & 0.5\%  & 0\%  & 0\%  & 0.08\%  & 0\%  & 0.1\%  & 0.18\%  \\
		
		& p & \textbf{92.4\%}   & \textbf{92.1\%}   & \textbf{86.8\%}    &  \textbf{43.9\%}       & 22.8\%  & 19\% & 10.8\%  & 53.5\%  & 67\%  & 0\%  & 0\%  & 80\%  & 0\%  & 34.1\%  & 47.4\%  \\
		
		& s & 0.03\% & 0.07\%  & 0.09\%     &  0.07\%    & 0.14\%   & 0.2\% & 0.06\%  & 0.12\%  & 0.18\%  & 0.002\%  & 0.004\%  & 0.03\%  & 0.05\%  & 0.08\%  & 0.1\%  \\
		\hline\hline
		
		\multirow{3}{*}{\hspace*{-1mm}2h5nd} & n 
		& 6.8\%   & 6.8\%   & 6.8\%    &  0\%    & 0\%   & 0\% & 0\%    & 0\%   & 0\%   & 0\%    & 0\%   & 0\%   & 0\%    & 0\%   & 0\%   \\
		
		& p & \textbf{94.1\%}   & \textbf{83.4\%}   & \textbf{71.4\%}    & 0\%      & 0\%  & 0\% & 0\%      & 0\%  & 0\%  & 0\%  & 0\%  & 0\%  & 0\%  & 0\%  & 0\%  \\
		
		& s & 0.028\% & 0.029\%  & 0.034\%     &  0.07\%    & 0.14\%   & 0.2\% & 0.06\%  & 0.11\%  & 0.17\%  & 0.065\%  & 0.075\%  & 0.077\%  & 0.03\%  & 0.07\%  & 0.09\%  \\
		\hline\hline
		
		\multirow{3}{*}{\hspace*{-1mm}1isua} & n 
		& 0.021\%   & 0.043\%   & 0.064\%    &  0.06\%    & 0.13\%   & 0.56\% & 0.02\%  & 0.11\%  & 0.11\%  & 0\%  & 0\%  & 0\%  & 0.02\%  & 0.17\%  & 0.17\%  \\
		
		& p & \textbf{17.5\%}   & 16.8\%   & 16.4\%    &  8.1\%      & 8.2\%  & \textbf{24.3\%} & 3.4\%   & 8.1\%   & 5.6\%   & 0\%   & 0\%   & 0\%   & 7.1\%   & \textbf{29.6\%}   & 16.7\%   \\
		
		& s & 0.01\% & 0.02\%  & 0.03\%     & 0.062\%    & 0.12\%   & 0.18\% & 0.05\% & 0.1\% & 0.15\% & 0.003\% & 0.013\% & 0.015\% & 0.023\% & 0.045\% & 0.08\% \\
		\hline\hline
		
		\multirow{3}{*}{\hspace*{-1mm}1cc5} & n 
		& 0.16\%   & 0.16\%   & 0.16\%    &  0\%    & 0.03\%   & 0.08\% & 0.6\%  & 0.65\%  & 0.97\%  & 0\%  & 0\%  & 0\%  & 0\%  & 0\%  & 0\%  \\
		
		& p & 50\%   & \textbf{42.7\%}   & \textbf{36.5\%}    & 0\%      & 1.6\%  & 3.1\% & \textbf{59.5\%} & 33.8\%  & 34.3\%  & 0\%  & 0\%  & 0\%  & 0\%  & 0\%  & 0\%  \\
		
		& s & 0.022\% & 0.026\%  & 0.03\%     &  0.05\%    & 0.11\%   & 0.17\% & 0.07\%  & 0.13\%  & 0.19\%  & 0.005\%  & 0.007\%  & 0.01\%  & 0.035\%  & 0.06\%  & 0.09\%  \\
		\hline\hline
		
		\multirow{3}{*}{\hspace*{-1mm}1aly} & n 
		& 4.12\%   & 5.2\%   & 6.2\%    & 0\%    & 0\%   & 0\% & 0\% & 0\% & 0\% & 0\% & 0\% & 0\% & 0\% & 0\% & 0\% \\
		
		& p & \textbf{42.6\%}   & \textbf{42\%}   & \textbf{41.7\%}    &  0\%   & 0\%  & 0\% & 0\%   & 0\%  & 0\%  & 0\%  & 0\%   & 0\%  & 0\%  & 0\%  & 0\%  \\
		
		& s & 0.035\% & 0.044\%  & 0.054\%     & 0.056\%    & 0.11\%   & 0.17\% & 0.051\%  & 0.11\%  & 0.16\%  & 0.01\%  & 0.024\%  & 0.026\%  & 0.025\%  & 0.055\%  & 0.081\%  \\
		\bottomrule
		
	\end{tabular}
	\label{tab:Comparison}
\end{table*}

\subsection{Quantitative Comparison of Decoy Selection Strategies}
\begin{figure*}[!t]
	\centering
	\includegraphics[width=0.95\textwidth]{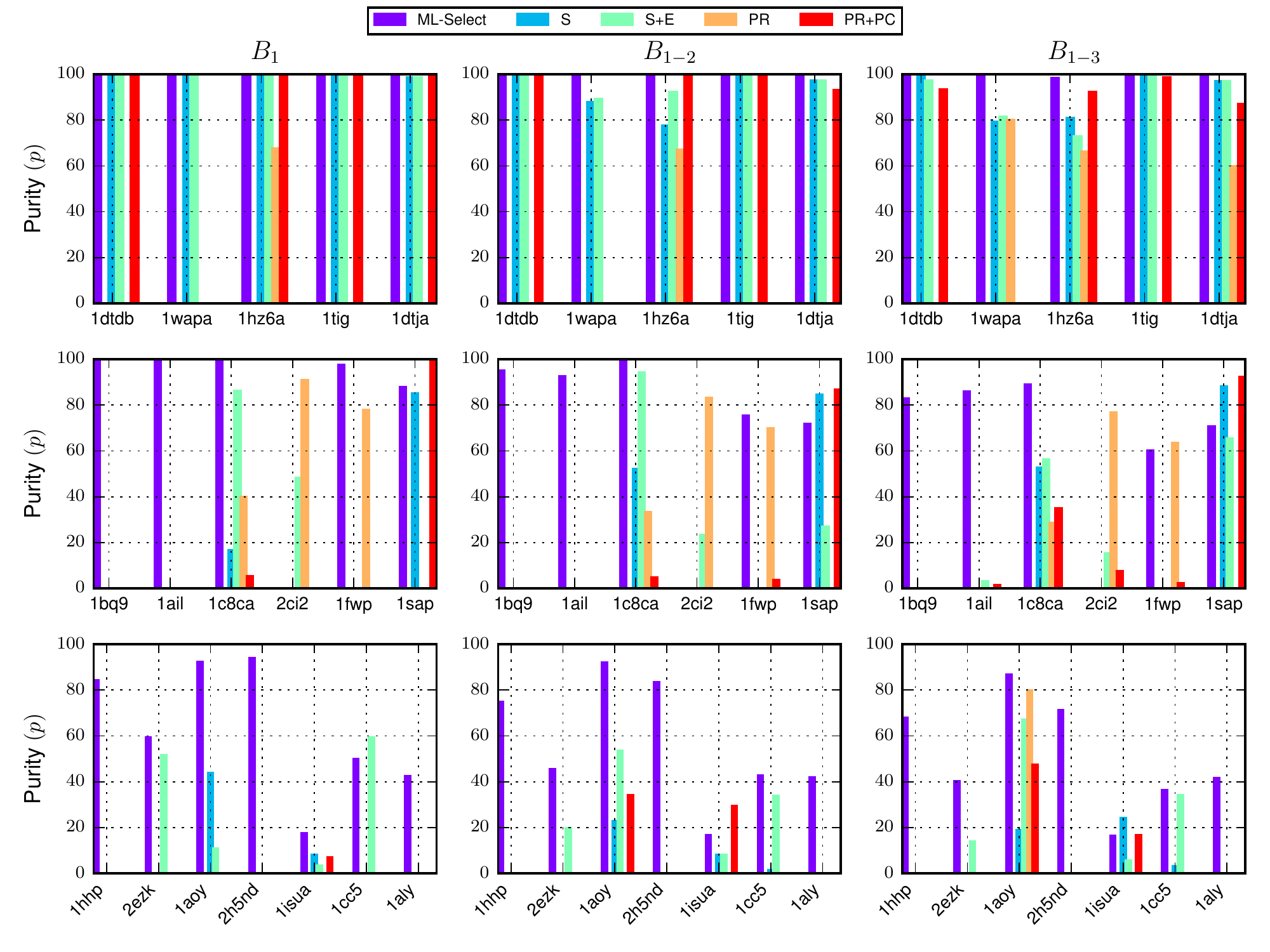}
	\caption{Comparison of the five selection strategies ML-Select, Size (S), Size+Energy (S+E), Pareto-Rank (PR), and Pareto-Rank+Count (PR+PC), in terms of the $p$ metric, for the easy, medium, and hard test cases. The top row shows the results for easy cases, second row is for the medium cases, and the bottom row shows the results for the hard cases. Metric $p$, purity, measures the percentage of near-native decoys in the $x$ selected basins while penalizing the basins by the extent of false positive presence. Results are shown for $x \in \{1,3\}$.}\label{fig:comparison}
\end{figure*}

Table~\ref{tab:Comparison} compares ML-Select with four basin-based decoy selection strategies proposed in~\cite{akhter2018extraction} on the easy, medium, and hard test cases. The comparison focuses on $p$ metric over $B_{1-x}$ groups of decoys where $x$ varies from $1$ to $3$. The results with respect to $n$ metric and the size ($s$) of each $B_{1-x}$ are also shown. Empirical evaluation conducted in~\cite{akhter2018extraction} shows that the four existing selection methods outperform a clustering-based decoy selection strategy. Figure~\ref{fig:comparison} compares the five selection strategies in terms of $p$ metric. The $x$-axis shows the test cases while $y$-axis tracks the purity ($p$) achieved by each method. The bold font indicates the best result among all the experimental methods. 

The purity of the top basin for all five selection strategies (except for PR, which performs much worse than others) are comparable for the easy cases (\textsl{1dtdb}, \textsl{1wapa}, \textsl{1hz6a}, \textsl{tig}, and \textsl{1dtja}). However, the purity diminishes as more basins are added to the selection for the four existing selection strategies (Size, Size+Energy, PR, PR+PC). For instance,  ML-Select scores more than $98\%$ for the top $3$ basins ($B_{1-3}$) for all the easy test cases, whereas {\em Basin-Size} can achieve only $79.3\%$ for \textsl{1wapa}, {\em Basin-Size+energy} can provide only $73\%$ purity for \textsl{1hz6a}, and {\em PR+PC} achieves $0\%$ purity for \textsl{1wapa}.

For the medium-difficulty cases, the purity improvements resulted from ML-Select are prominent. ML-Select outperforms the four existing selection strategies in $4$ out of $6$ cases for $B_{1-x}$, where $x \in [1-3]$. For instance, ML-Select achieves a maximum of $100\%$ and a minimum of $83\%$ purity for \textsl{1bq9} and \textsl{1ail}, whereas the remaining four methods achieve a minimum of $0\%$ purity and a maximum of $3\%$ purity. 

The hard cases present the most challenging decoy ensembles. Even for these challenging decoy sets, ML-Select significantly outperforms the four existing selection strategies in $5$ out of $7$ test cases (\textsl{1hhp}, \textsl{2ezk}, \textsl{1aoy}, \textsl{2h5nd}, and \textsl{1aly}) for all sizes of basin selections (i.e., $B_{1-x}$, $x \in [1-3]$). For two other cases (\textsl{1isua} and \textsl{1cc5}), ML-Select performs better for the top basin for \textsl{1isua}, and for \textsl{1cc5} when $x \in [2,3]$. For instance, for the most difficult test case \textsl{1aly}, ML-Select obtains about $42\%$ purity whereas the four other methods fail to provide a single true positive ($0\%$ purity).

Table 3 compares ML-Select with MUFOLD-CL on the easy, medium, and hard test cases. For all cases, the top three clusters are fairly large, which lowers purity. For instance, the smallest of the top clusters (on \textsl{1wapa}) contains $39\%$  of all the decoys in the decoy set of size $68,000$. The near-native presence in this decoy set is only $0.005\%$. As a result, despite containing $39.4\%$ near-natives, abundant non-natives populating the top cluster lowers its purity. In contrast, ML-Select is more precise; it selects basins of much smaller size that consist of mostly near-natives, resulting in much higher purity.

Figure~\ref{fig:comparison} shows that ML-Select offers reasonably good performance for a variety of test cases, which is not the case with the basin-based strategies. For instance, PR performs quite well for \textsl{1c8ca} and \textsl{2ci2} for $B_1$, but it fails miserably for \textsl{1bq9}, \textsl{1ail}, and \textsl{1sap}. As a result, one cannot rely on this selection strategy in achieving good purity over a new test case. Contrarily, ML-Select guarantees reasonably good purity over all the test cases (except for one test case, \textsl{2ci2}). Hence, ML-Select stands out as a more reliable decoy selection strategy than the four existing selection methods. 

\begin{figure*}[!t]
	\centering
	\includegraphics[width=0.95\textwidth]{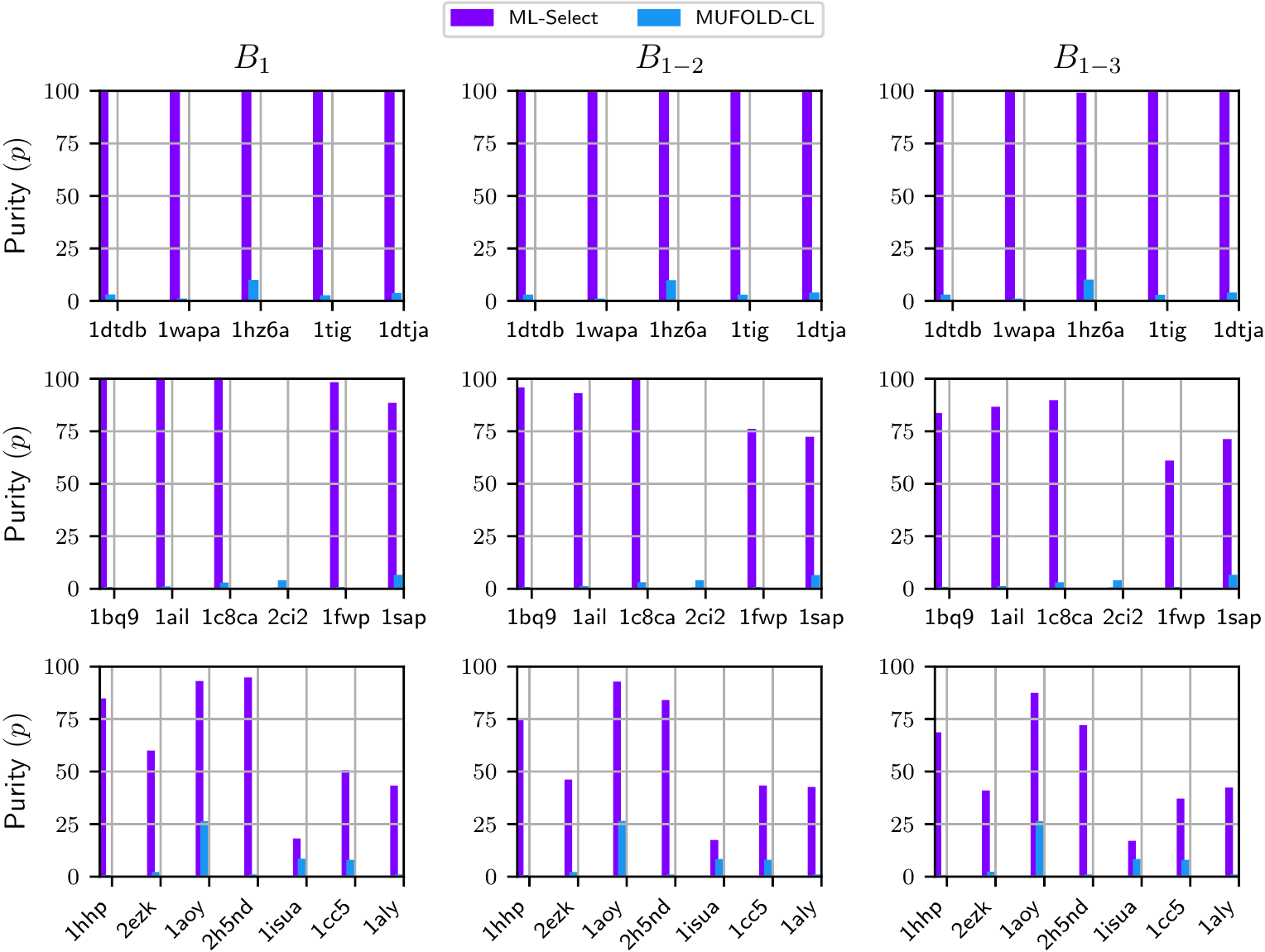}
	\caption{Comparison of ML-Select and MUFOLD-CL, in terms of the $p$ metric, for the easy, medium, and hard test cases. The top row shows the results for easy cases, second row is for the medium cases, and the bottom row shows the results for the hard cases. Metric $p$, purity, measures the percentage of near-native decoys in the $x$ selected basins while penalizing the basins by the extent of false positive presence. Results are shown for $x \in \{1,3\}$.}
	\label{fig:comparison_with_mufold}
\end{figure*}

Figure~\ref{fig:comparison_with_mufold} shows that ML-Select performs much better than MUFOLD-CL in terms of the purity metric. However, MUFOLD-CL has been able to provide some near-natives for the medium-difficulty protein \textsl{2ci2} on which ML-Select obtains $0\%$ purity. However, MUFOLD-CL's performance in terms of purity is low, as well. This is due to the much bigger cluster size and the scarcity of near-natives in the decoy sets.

\begin{table}[htp] 
	\centering
	\caption{Statistical significance of five methods over eighteen test-cases determined through Friedman tests with Hommel's post-hoc analysis at $\alpha$=$0.05$. The best method is marked with an asterisk (*), while the boldface presents the significance of the respective method when compared with the best method.}
	\begin{tabular}{c|l|p{3em}|c|c}
		\toprule
		Top &\multirow{2}{*}{Method} & Average & $p$ &$p$\\
		Basins & & Rank & value & Hommel\\
		\midrule
		\multirow{5}{*}{$B_1$}& \textbf{PR} & \textbf{3.889} & \textbf{2.101E-6} & \textbf{0.0125} \\
		&\textbf{Basin-Size} & \textbf{3.306} & \textbf{2.76E-4} & \textbf{0.0167} \\
		&\textbf{PR+PC}& \textbf{3.306} & \textbf{2.76E-4} & \textbf{0.025} \\
		&\textbf{Basin-Size+Energy} & \textbf{3.11} & \textbf{0.001} & \textbf{0.05} \\
		&ML-select$^*$& 1.389 & - & - \\
		\midrule
		\multirow{5}{*}{$B_{1-2}$}& \textbf{PR} & \textbf{4.028} & \textbf{5.53E-7} & \textbf{0.0125} \\
		&\textbf{Basin-Size} & \textbf{3.417} & \textbf{1.19E-4} & \textbf{0.0167} \\
		&\textbf{PR+PC}& \textbf{3.139} & \textbf{8.99E-4} & \textbf{0.025} \\
		&\textbf{Basin-Size+Energy} & \textbf{3.028} & \textbf{0.002} & \textbf{0.05} \\
		&ML-Select$^*$& 1.389 & - & - \\
		\midrule
		\multirow{5}{*}{$B_{1-3}$}& \textbf{PR} & \textbf{3.833} & \textbf{7.47E-6} & \textbf{0.0125} \\
		&\textbf{Basin-Size} & \textbf{3.444} & \textbf{1.83E-4} & \textbf{0.0167} \\
		&\textbf{PR+PC}& \textbf{3.306} & \textbf{5.04E-4} & \textbf{0.025} \\
		&\textbf{Basin-Size+Energy} & \textbf{2.944} & \textbf{0.005} & \textbf{0.05} \\
		&ML-Select$^*$& 1.472 & - & - \\
		\bottomrule
	\end{tabular}
	\label{tab:stats} 
\end{table}

Table~\ref{tab:stats} shows the Friedman statistical tests with Hommel’s post-hoc~\cite{garcia2008extension} analysis in predicting the purity of the basins. The statistical tests are performed on all the five different experimental methods on all the eighteen test case proteins at $\alpha$ = $0.05$. The first column indicates the number of basins under consideration in the prediction of purity. The second column shows the methods, while the third column presents the average rank calculated from the Friedman's test~\cite{demvsar2006stat}, which rejects the null hypothesis. Upon the rejection of the null hypothesis, Hommel's post-hoc analysis helps to determine the statistical significance of the new technique (ML-Select) when compared to that of the existing methods. The fourth and the fifth columns show the $p$-value and Hommel’s critical value respectively. The lowest average rank shows the best (ML-Select) method, and is marked with an asterisk (*). A method is said to be significantly different from the best method if the $p$-value of the corresponding method is less than that of the $p$-Hommel at $\alpha$ = 0.05, is in boldface. Overall, for all the three different basin sizes, {\em ML-Select} is the best. Therefore, ML-Select significantly outperforms the existing basin-based selection strategies.

\subsection{Effect of $\mathit{dist\_thresh}$ on Performance}
We varied the $\mathit{dist\_thresh}$ parameter in the second phase to monitor any performance deviations in ML-Select. Here we summarize our findings. The improvement in the purity of the selected basins is insignificant when we alter the pre-defined distance threshold, $\mathit{dist\_thresh} \pm \tau$, where $\tau \in {10\%, 20\%, 25\%}$. In $15$ out of $18$ test cases, the purity varied, however, when $\mathit{dist\_thresh}$ is increased by $20\%$, we see an insignificant improvement. For example, the purity of the top $3$ basins for \textsl{1bq9} increases from $83\%$ to $94.6\%$ when the $\mathit{dist\_thresh}$ is raised by $20\%$. For all the remaining test cases, the improvement in the purity is insignificant. Overall, altering the distance threshold by a factor has insignificant impact in predicting the purity.

\section{Discussion}
\label{sec:Discussion}
The results presented in this paper suggest that energy landscape probed by a template-free protein structure  prediction method can be leveraged for decoy selection and warrants further investigation. In particular, energy is often ignored in favor of structural similarity in clustering-based decoy selection strategies. The work presented in this paper has demonstrated that energy, when utilized in the context of energy landscape, can be successfully employed to identify near-native decoys from a decoy ensemble. 

Observation on results from clustering-based selection methods show that these methods fail to identify exceptionally good decoys for sparsely distributed decoy ensembles. Since a clear consensus is often not available as near-native decoys are usually scarce and far away from the rest of the decoys, consensus-based methods such as clustering-based selections struggle to yield good performance for such challenging datasets. As shown in this paper, basins in energy landscape can improve decoy selection performance. In particular, supervised learning methods applied to basins extracted from an energy landscape can not only provide better decoy selection performance, but also prove resilient against sparsely distributed decoy ensembles.

Specifically, this paper presents a novel decoy selection method, ML-Select, that employs a supervised machine learning method to identify  basins comprising mostly near-native decoys. ML-Select utilizes both energy- and graph-based characteristics of basins to successfully select near-native basins even for the challenging datasets consisting of only a few near-natives. Results presented in this paper also show that ML-Select is able to provide good performance for varied test cases irrespective of the difficulty level of the decoy ensemble.

Although ML-Select shows promise in decoy selection in template-free protein structure prediction, further investigation is warranted to address the current limitations. For instance, while ML-Select is able to provide a  good-quality basin, this method does not assess the quality of individual decoys in the selected basin. However, the selected basin offers an informative set from which the best decoy(s) can be identified with the help of further ranking and more investigation. Further work will concentrate on utilizing decoy characteristics to incorporate an weighting scheme for identifying the best decoy(s) from a decoy ensemble. The line of inquiry pursued in this paper demonstrates a promising direction for advancing decoy selection research.

\section{Conclusion}
\label{sec:Conclusions}

We proposed a novel machine learning strategy, ML-Select, in purifying the basins generated from the energy landscapes. Our experimental results indicate the utility of basins in the energy landscape probed by a template-free structure prediction method for automatic decoy selection. The model has been evaluated in terms of purity (favors lower false-positives and higher true-positives) and compared against four existing basin-based decoy selection strategies that perform better than a cluster-based selection strategy. We showed that ML-Select performs significantly better than all the four basin-based selection strategies. Moreover, the performance of ML-Select is highly reliable, unlike the inconsistent dominance of basin-based methods over the cluster-based method. Finally, we validate the use of machine learning techniques in decoy selection, while suggesting further research in this direction for advancing the state of decoy selection. In the future, we would like to investigate the use of other machine learning strategies and/or heuristics (similar to~\cite{chennupati2014efficiency,Chennupati:2015}) that initially predict the difficulty of a protein and use an ensemble of algorithms in predicting the purity of the basins for the respective class of proteins.


\bigskip

\section{List of abbreviations}

ML -- Machine Learning\\
PDB -- Protein Data Bank\\
RMSD -- Root Mean Squared Deviation\\
PR -- Pareto Rank\\
PC -- Pareto Count\\
SBL -- Structural Bioinformatics Library

\section{Acknowledgements}

Computations were run on Darwin, a research computing heterogeneous cluster
(URL: https://darwin.lanl.gov).

\section{Funding}

The research was supported by Los Alamos National Laboratory (LANL) LDRD ER grant (20160317ER). Parts of this research used resources provided by the Los Alamos National Laboratory Institutional Computing Program, which is supported by the U.S. Department of Energy National Nuclear Security Administration under Contract No. DE-AC52-06NA25396. This work is also supported in part by the National Science Foundation Grant No. 1900061. This material is additionally based upon work supported by (while serving at) the National Science Foundation.  Any opinion, findings, and conclusions or recommendations expressed in this material are those of the author(s) and do not necessarily reflect the views of the National Science Foundation. Publication costs are funded by the National Science Foundation. The funder has no role in the research and writing of the paper.

\section{Availability of data and materials}

All software and data are available upon demand.

\section{Authors' contributions}

NA drafted the manuscript. NA, GC, DH, and AS revised the manuscript. NA designed and executed the experiments, while GC, DH, and AS supervised the design and analysis of methods. NA implemented majority of the code, while GC implemented the graph features. NA, GC, DH and AS conceptualized the methods. All authors provided critical feedback on the manuscript, read and approved the final manuscript.

\section{Ethics approval and consent to participate}

Not applicable.

\section{Consent to publish}

Not applicable.

\section{Competing interests}

The authors declare that they have no competing interests.

\newpage
\clearpage

\bibliographystyle{abbrv}

\end{document}